\documentclass[aps,prb,preprintgroupedaddress,eqsecnum,draft,showpacs,intlimits,amsmath,amssymb]{revtex4}
\usepackage{bm}
\usepackage[final]{graphicx}
\usepackage{longtable}
\newcommand{\be}{\begin{equation}}
\newcommand{\ee}{\end{equation}}
\newcommand{\bea}{\begin{eqnarray}}
\newcommand{\eea}{\end{eqnarray}}

\begin{document}

\title{Critical adsorption of polymers in a medium with long-range correlated quenched disorder.}

\author{Z. Usatenko$^{1,2}$, A.Ciach$^2$}

\affiliation{$^{1}$Institute for Condensed Matter Physics,
National Academy of Sciences of Ukraine, 1~Svientsitskii Str.,
UA--79011 Lviv, Ukraine} \email{pylyp@ph.icmp.lviv.ua}
\affiliation{$^{2}$Institute of Physical Chemistry of the Polish
Academy of Sciences, Warsaw, 01-224, Poland}

%\date{\today}

%\vspace{0.1cm}

\begin{abstract}

We investigated the influence of short- and long-range correlated
quenched disorder introduced into the medium on the process of
adsorption of long-flexible polymer chains on a wall by using the
field theoretical approach in $d=4-\epsilon$ and directly in $d=3$
dimensions up to one-loop order for the semi-infinite $|\phi^4|$
m-vector model (in the limit $m\to 0$) with a boundary. This
allows us to describe the critical behaviour of long-flexible
polymer chains in the vicinity of the surface and to obtain the
whole set of surface critical exponents at the special surface
transition ($c=c_{ads}$), which separates the nonadsorbed region
$c>c_{ads}$ from the adsorbed  one $c<c_{ads}$. In the case of
very strong correlation of the disorder we obtained that a
 polymer collapses and then adsorbs on a wall.
The obtained results indicate that for the systems with
long-range-correlated quenched disorder the new set of surface
critical exponents arises and show that polymer chains in solution
with that type of disorder adsorb on a wall stronger than polymer
chains in pure medium.

\end{abstract}

\maketitle
\renewcommand{\theequation}{\arabic{equation}}
 The critical behavior of the real systems with
different kinds of disorder is of considerable theoretical and
experimental interest. In common, the real physical systems are
usually characterized by the presence of different kinds of defect
and impurities  that may be localized inside the bulk or at the
boundary. The present work is connected with investigation of
adsorption phenomena of long-flexible polymer chains in disordered
media with correlated structural defects. The main question is:
does a small amount of long-range correlated quenched disorder
introduced into the medium induce changes to the universal
properties of long-flexible polymer chains at the adsorption on a
wall?

As it is known \cite{Cloizeaux}, long flexible polymer chains in a good
solvent are perfectly described by a model of self-avoiding walks (SAW) on
a regular lattice. Their scaling properties in the limit of infinite
number of steps may be derived by a formal $m \to 0$ limit of the $O(m)$
vector model at its critical point \cite{deGennes}. Thus for the average
square end-to-end distance $R$ and the number of configurations $Z_{N}$ of
a SAW with $N$ steps and one end fixed or with both ends fixed on a
regular lattice in the asymptotic limit $N\to \infty$ the following
relations take place
\be
<R^2>\sim N^{2\nu},\quad\quad\quad Z_{N}\sim
q^{N}N^{\gamma-1},\quad\quad\quad Z_{N}(r)\sim
q^{N}N^{-(2-\alpha)},\label{RZ} \ee
where $\nu$ and $\gamma$ are the
univeral correlation length and susceptebility exponents for the $m=0$
model, $d$ is the space dimensionality, $q$ is a nonuniversal fugacity and
$r^{2}=({\vec r}_{A}-{\vec r}_{B})^{2}$. What happens when disorder will
be introduced into the system? As it is known the infinite systems are
described by the Landau-Ginzburg-Wilson Hamiltonian \cite{deGennes}

\be
H = \int_{V} d^{d}x [\frac{1}{2}
\mid \nabla\vec{\phi} \mid ^{2} +
\frac{1}{2} (m_{0}^{2}+\delta\tau({\bf{x}}))\mid \vec{\phi} \mid^{2}
+\frac{1}{4!} v_{0} (\vec{\phi}^2)^{2}],\label{1}
\ee
where $\vec{\phi}(x)$ is an $m$-vector field with the
components $\phi_{i}(x)$, $i=1,...,m$. The inhomogeneities in the system
cause fluctuations in the local transition temperature. Here $m_{0}^2$ is
the "bare mass", which can be interpreted in this case as
chemical potentials for monomers in the bulk, $\delta\tau({\bf{x}})$
represents the quenched random-temperature disorder, with
$<\delta\tau({\bf{x}})>=0$ and
\be
\frac{1}{8}<\delta\tau({\bf{x}})\delta\tau({\bf{x}}')>=g(\mid
{\bf{x}}\mid), \label{2} \ee
where angular brackets $<...>$ denote configurational averaging over
quenched disorder. The pair correlation function $g(\mid {\bf{x}}\mid)$ is
assumed to fall off with the distance as
\be
g(\mid {\bf{x}}\mid) \sim
\frac{1}{x^a}\label{2}
\ee
for large ${\bf{x}}=({\bf{r}},z)$, where $a$ is a constant and
$x=\mid \bf{x}\mid$.

The Fourier-transform ${\tilde{g}}(k)$ of $g(x)$ for small $k$ is

\be
{\tilde g}(k) \sim u_{0}+w_{0} \mid k\mid^{a-d}.\label{3}
\ee
This corresponds to the so-called long-range-correlated
"random-temperature" disorder.
In the case of random uncorrelated pointlike (or short-range-correlated)
disorder the site-occupation correlation function is $g(x) \sim \delta(x)$
and its Fourier-transform
\be
{\tilde g}(k) \sim u_{0}.\label{4}
\ee

Applying the replica method in order to average the free energy over
different configurations of quenched disorder, it is possible to construct
an effective Hamiltonian of the $|\phi^4|$ $m$ -vector model with a
long-range-correlated disorder
\bea
H_{eff} & = & \sum_{\alpha=1}^{n}\int_{V} d^{d}x [\frac{1}{2}
\mid \nabla\vec{\phi}_{\alpha} \mid ^{2} +
\frac{1}{2} m_{0}^{2}\vec{\phi}_{\alpha}^{2}
+\frac{1}{4!} v_{0}
(\vec{\phi}_{\alpha}^2)^{2}\nonumber\\
&& -\sum_{\alpha,\beta=1}^{n}\int
d^{d}x_{1} d^{d}x_{1}^{'}g(\mid
x_{1}-x_{1}^{'}\mid)\vec{\phi}_{\alpha}^{2}
(x_{1})\vec{\phi}_{\beta}^{2}(x_{1}^{'}).\label{5}
\eea

We are interested in investigation of the polymer limit $m\to 0$ of such
kind of model which can be interpreted as a model of long-flexible
polymer chains in a disordered medium. Here Greek indices denote
replicas, and the replica limit $n\to 0$ is implied.
If $a\geq d$, than $w_{0}$ term
irrelevant. This corresponds to {\it random uncorrelated pointlike
disorder} (or short-range-correlated random disorder).  As noticed by Kim
\cite{Kim}, in this case in the limit $m,n \to 0$ both $v_{0}$ and $u_{0}$
terms are of the same symmetry. It indicates that {\it a weak quenched
uncorrelated disorder is irrelevant for SAWs}\cite{Harris}. Thus, it is
imposible to "naively" apply the Harris criterion \cite{Harris74} to the
SAW problem, though the critical exponent $\alpha$ of a SAW on the $d=3$
dimensional pure lattice is positive ($\alpha(d=3)=0.235\pm 0.003$
\cite{Guida98}).

If $a<d$, the term $w_{0}k^{a-d}$ is relevant and {\it the
long-range-correlated disorder is relevant for SAWs} (see
\cite{Holovatch01}).

The presence of a surface or a wall leads to the appearance of
additional problems in analysis of the critical behavior. The
investigation of adsorption phenomena of long-flexible polymers on a hard
wall was a subject of a series of works (for the sake of brevity we notice
only few of them \cite{deGennes,EKB82,Eisenriegler}).
It should be mentioned that in the case of systems with a surface, the
different kinds of defects and impurities may be
 localized inside the bulk or at the boundary.

As was found in \cite{DN89}, introducing into the system {\it
short-range correlated random quenched surface disorder} is irrelevent for
critical behavior, but long-range correlated quenched surface disorder with
$
g(r)\sim \frac{1}{r^{a}}
$
and with the corresponding Fourier transform
\be
{\tilde g}(p)\sim u_{0}+w_{0}p^{a-d+1}\label{6}
\ee
can be relevant, if $a<d-1$ and is irrelevant if $a\geq d-1$.

Our present work is connected with the investigation of the
adsorption phenomena on a planar wall of long flexible polymer chains
inserted into disordered medium with long-range-correlated quenched
disorder. The solution of polymer chains in this semi-infinite
space is sufficiently dilute, so that interchain interactions can be
neglected and it is sufficient to consider surface effects for
configurations of a single chain.

We pay particular attention to the investigation of
special surface transition ($c=c_{ads}$), which separates the nonadsorbed
region $c>c_{ads}$ from the adsorbed  one $c<c_{ads}$. In the latter two
regions, the chemical potential $c$ for monomers on the surface does not
stay fixed but tends to $+\infty$ and $-\infty$, respectively
\cite{EKB82,Eisenriegler}.

The effective Hamiltonian of the semi-infinite $|\phi^4|$ $m$-vector model
with long-range-correlated disorder in this case is

\bea
H_{eff} & = & \sum_{\alpha=1}^{n}\int_{V} d^{d}x [\frac{1}{2}
\mid \nabla\vec{\phi}_{\alpha} \mid ^{2} +
\frac{1}{2} m_{0}^{2}\vec{\phi}_{\alpha}^{2}
+\frac{1}{4!} v_{0}
(\vec{\phi}_{\alpha}^2)^{2}\nonumber\\
&& -\sum_{\alpha,\beta=1}^{n}\int
d^{d}x_{1} d^{d}x_{1}^{'}\bar{g}(\mid
x_{1}-x_{1}^{'}\mid)\vec{\phi}_{\alpha}^{2}
(x_{1})\vec{\phi}_{\beta}^{2}(x_{1}^{'})+\frac{c_{0}}{2}\sum_{\alpha=1}^{n}
\int_{\partial V}d^{d-1}r \vec{\phi}_{\alpha}^{2}({\bf{r}},z=0).\label{9}
\eea

In accordance with the fact that we have to deal with semi-infinite
geometry $({\bf x}=({\bf r},z\geq 0))$, only parallel Fourier
transformations in $d-1$ dimensions will be performed. The parallel
Fourier transform ${\tilde{g}}(q,z)$ of (\ref{2}) is
\be
{\tilde{g}}(q,z)\sim
w_{0}\frac{2^{\frac{a-d+1}{2}}}{\Gamma[\frac{d-a}{2}]\sqrt{\pi}}
q^{\frac{a-d+1}{2}}z^{\frac{d-a-1}{2}} K_{\frac{a-d+1}{2}(q z)},\label{7}
\ee
where $K_{\frac{a-d+1}{2}(q z)}$ is the modified Bessel function and
$q=\mid{\bf q}\mid$, where ${\bf q}$ is $d-1$ dimensional vector. In the
case of small $q$ and $z$ we obtain the relation
\be
{\tilde g}(q,z)\sim u_{0}+w_{0}q^{a-d+1}+w_{0}^{'}z^{d-a-1},\label{8}
\ee
which agrees with the predictions obtained in \cite{DN89} (see (\ref{6})
in the case $z=0$).
In the general case of arbitrary $z$ (from $z=0$ on the wall to
$z\to \infty$) we must take into account the Fourier transform
${\tilde{g}}(q,z)$ of the form (\ref{7}).

The corresponding correlation function of the model (\ref{9}), which
involves $N$ fields $\phi({\bf{x}}_{i})$ at distinct points
${\bf{x}}_{i}(1\leq i \leq N)$ in the bulk and $M$ fields
$\phi({\bf{r}}_{j},z=0)\equiv \phi_{s}({\bf{r}}_{j})$ at distinct
points on the wall with parallel coordinates ${\bf{r}}_{j}(1\leq j \leq
M)$, has the form

\be
G^{(N,M)}(\{{\bf x}_{i}\},\{{\bf{r_{j}}}\}) = < \prod_{i=1}^{N}
\phi({\bf x}_{i})\prod_{j=1}^{M}\phi_{s}({\bf r}_{j})>. \label{10}
\ee

The corresponding full free propagator in the mixed ${\bf{p}} z$
representation is given by

\be
G({{\bf{p}}},z,z') = \frac{1}{2\kappa_{0}} \left[ e^{-\kappa_{0}|z-z'|} -
\frac{c_{0}-\kappa_{0}}{c_{0}+\kappa_{0}} e^{-\kappa_{0}(z+z')}
\right],\label{11}
\ee
where
$
\kappa_{0}=\sqrt{p^{2}+m_{0}^{2}}
$
with $p$ being the value of the parallel momentum ${\bf p}$ associated
with the $d-1$ translationally invariant directions in the system.

There are two possible cases: a) when two ends of polymer are attached to the
wall (in such a case we have to deal with a calculation of two point
correlation function $G^{(0,2)}(r,z=0;r',z'=0)$), and b) when one end of
the polymer is in the bulk and the other one is attached to the wall
$(G^{(1,1)}(x;r',z'=0)$).
In order to obtain the universal surface critical exponents characterizing
the adsorption of long-flexible polymer chains inserted into the medium
with long-range-correlated quenched disorder on the wall, it is sufficient
to consider the correlation function of two
surface fields $G^{(0,2)}(r,z=0;r',z'=0)$. The universal surface critical
exponents for such systems depend on the dimensionality of space $d$,
the value of order parameter components $m (m\to 0)$ and the range of the
correlations, i.e. the value of  $a$.

As it is known, in the theory of
semi-infinite systems the bulk field $\phi({\bf x})$ and the surface
field $\phi_{s}({\bf r})$ should be reparameterized by different uv-finite
renormalization factors $Z_{\phi}(u,v,w)$ and $Z_{1}(u,v,w)$,
$$ \phi(x) =
Z_{\phi}^{1/2}\phi_{R}(x)\quad\quad\quad {\rm and} \quad\quad\quad
 \phi_{s}(r)=Z_{\phi}^{1/2}Z_{1}^{1/2}\phi_{s,R}(r). $$
The renormalized correlation function involving $N$ bulk and $M$
surface fields can be written as
\be
G_{R}^{(N,M)} ({\bf{p}} ; m,u,v,w,c)=Z_{\phi}^{-(N+M)/2} Z_{1}^{-M/2}
G^{(N,M)} ({\bf{p}} ; m_{0},u_{0},v_{0},w_{0},c_{0}).\label{12}
\ee

It should be mentioned that the typical bulk short-distance singularities,
which are present in the correlation function $G^{(0,2)}$, can be
subtracted after performing the mass renormalization. For distinguished
the parallel and perpendicular directions we obtain:

\be
m_{0}^{2}=m^{2}-t_{1}^{(0)} I_{1}(m^{2})+
t_{2}^{(0)} I_{2}(m^{2}),\label{14}
\ee
where
\be
t_{1}^{(0)}=\frac{1}{3}(v_{0}-u_{0}-\frac{w_{0}m^{a-d}}{cos[\pi(a-d)/2]}),
\quad\quad\quad
t_{2}^{(0)}=\frac{w_{0}}{3\sqrt{\pi}}\frac{\Gamma(\frac{d-a-1}{2})}{\Gamma(\frac{d-a}{2})},\label{14a}
\ee and
\be
I_{1}(m^2)=\frac{1}{(2\pi)^{d-1}}\int\frac{d^{d-1}q}{2\kappa_{q}}\label{I1}
\ee with $\kappa_{q}=\sqrt{q^2+m^2}$ and
$$I_{2}=\frac{1}{(2\pi)^{d-1}}\int d^{d-1}q \frac{\mid q\mid^{a-d+1}
{}_2F_{1}[\frac{1}{2},1,\frac{3+a-d}{2},\frac{q^2}{\kappa_{q}^2}]}{2
\kappa_{q}^2}.$$
Acording to the above mentioned notations, we have in the
effective Hamiltonian only two coupling constants, $V_{0}=v_{0}-u_{0}$ and
$w_{0}$ (we keep notation $v_{0}$ for $V_{0}$).

The renormalized coupling constants $v$, $w$
are fixed via the standard normalization conditions of the
infinite-volume theory \cite{Holovatch01}
\bea
m^{4-d}v&=&\Gamma_{b,R,v}^{(4)}(\{q\};m^2,v,w)|_{q=0},\nonumber\\
m^{4-a}w&=&\Gamma_{b,R,w}^{(4)}(\{q\};m^2,v,w)|_{q=0},\label{14b}
\eea

where $\Gamma_{b,R,v}^{(4)}$ and $\Gamma_{b,R,w}^{(4)}$ are the $v$
and $w$ term symmetry contributions to the four-point vertex function.
To the present accuracy of calculation at one-loop order, the vertex
renormalization gives: $v=v_{0}m^{d-4}$ and $w=w_{0}m^{a-4}$.

 In order to remove the short-distance singularities of
the correlation function $G^{(0,2)}$ located in the vicinity of the
surface, the surface-enhancement shift $\delta c$ is required. In accordance
with this, the new normalization condition should be introduced for the
definition of the surface-enhancement shift $\delta c$ and surface
renormalization factor $Z_{1}$. By analogy with magnetic systems
\cite{DSh98,Sh97,UH02}, the renormalized surface two-point correlation
function in our case is normalized in such a manner \cite{DSh98} that at
zero external momentum it should coincide with the lowest order
perturbation expansion of the surface susceptibility
$\chi_{\parallel}(p)=G^{(0,2)}(p)$ \be G^{(0,2)}
(p;m_{0},v_{0},w_{0},c_{0}) = \frac{1}{c_{0}+\sqrt{p^{2} + m_{0}^{2}}} +
O(v_{0},w_{0})\label{15} \ee
and its first derivatives with respect to $p^2$.
Thus we obtain the necessary surface normalization conditions

\be
G_{R}^{(0,2)}(0;m,v,w,c) = \frac{1}{m+c}\label{16}
\ee
and

\be
\left.\frac{\partial G_{R}^{(0,2)} (p;m,v,w,c)}{\partial p^{2}}
\right|_{p=0} = -
\frac{1}{2m(m+c)^{2}}. \label{17}
\ee

Equation (\ref{16}) defines the required surface-enhancement shift $\delta c$
and shows that the surface susceptibility
diverges at $m=c=0$. This point corresponds to the multicritical
point $(m_{0c}^{2},c_{0}^{sp})$ at which the special transition takes
place.

From the normalization condition of Eq. (\ref{17}) and
expression for the renormalized correlation function of Eq. (\ref{12}), we
can define the renormalization factor $Z_{\parallel} = Z_{1} Z_{\phi}$ by

\be
Z_{\parallel}^{-1} = \left. 2m \frac{\partial}{\partial
p^{2}}[G^{(0,2)} (p)]^{-1}\right|_ {p^2=0} = \lim_{p\to
0}{m\over p}{\partial\over\partial p} [G^{(0,2)}(p)]^{-1}.\label{18}
\ee

Asymptotically close to the critical point the two point renormalized
correlation functions $G_{R}^{(N,M)}$ satisfy the corresponding
homogeneous Callan-Symanzik equations \cite{Sh97,DSh98}
\begin{eqnarray}
&& \left[ m\frac{\partial}{\partial m}+\beta_{v}
(v,w)\frac{\partial}{\partial v}+ \beta_{w} (v,w)\frac{\partial}{\partial
w}+\frac{N+M}{2}\eta (v,w)\right.\nonumber\\
&& +\left.\frac{M}{2}\eta^{sp}_{1}(v,w)\right]
G^{(N,M)}_{R}(0;m,v,w,c)= 0,\label{19}
\end{eqnarray}

where the $\beta$-functions are
$\beta_{v}(v,w)=\left.m\frac{\partial}{\partial m}\right|_{L\!R}
v,\quad\beta_{w}(v,w)=\left.m\frac{\partial}{\partial
m}\right|_{L\!R} w $, the exponents $\eta$ and $\eta_{1}^{sp}$
are \be
\eta=m\!{\partial\over \partial m}
\left.\ln\!{Z_{\phi}}\,\right|_{L\!R},\quad\quad\quad
\eta_{1}^{sp}=m\!{\partial\over \partial m}
\left.\ln\!{Z_{1}}\,\right|_{L\!R},\label{19}
\ee
and where LR is the long-range fixed point.
The simple scaling dimensional analysis of $G_{R}^{(0,2)}$ and mass
dependence of $Z$ factors, defines the surface correlation
exponent $\eta_{\parallel}^{sp}$ via
\be
\eta_{\parallel}^{sp}=\eta_{1}^{sp}+\eta. \label{20}
\ee
From Eqs. (\ref{18}),(\ref{19}) and (\ref{20}), we obtain for the surface
correlation exponent $\eta_{\parallel}^{sp}$

\begin{eqnarray}
\eta_{\parallel}^{sp}&=&m\!{\partial\over \partial m}
\left.\ln\!{Z_{\parallel}}\,\right|_{L\!R}\nonumber\\
&=&\beta_v(v,w){\partial\ln Z_{\parallel}(v,w)\over \partial v}+
\left.\beta_w(v,w){\partial\ln Z_{\parallel}(v,w)\over \partial
w}\,\right|_{L\!R},\label{21}
\end{eqnarray}
where the one-loop pieces of the $\beta$ functions are \cite{Holovatch01}
\bea
\beta_{\bar
v}({\bar v},{\bar
w})&=&-{\bar{v}}+{\bar{v}}^2-(3f_{1}(a)-f_{2}(a)){\bar{v}}
{\bar{w}},\nonumber\\ \beta_{\bar w}({\bar v},{\bar w})&=&-(4-a){\bar
w}-(f_{1}(a)-f_{2}(a)){\bar w}^2+\frac{{\bar v} {\bar w}}{2}.\label{22}
\eea
The renormalized coupling constants $v$ and $w$ are normalized in a
standard fashion so that
$$
{\bar v}=\frac{4}{3}v I_{1},\quad\quad\quad {\bar w}=\frac{4}{3} w I_{1},
$$
and the integral $I_{1}$ (see [\ref{I1}]) in the case of $d=3$ is equal to
$1/8\pi$ and in the case of $d=4-\epsilon$ it is
$I_{1}=2^{-d}\pi^{-d/2}\Gamma (\epsilon/2)$ ( the case
$\epsilon=4-d$ and $\delta=4-a$ is presented in Appendix 1). The
coefficients $f_{i}(a)$ expressed via the one-loop integrals represented
in \cite{Holovatch01} and are \cite{Prudnikov}

\bea f_{1}(a)&=&\frac{(a-2)(a-4)}{2 \sin (\pi
a/2)},\nonumber\\ f_{2}(a)&=&\frac{(a-2)(a-3)(a-4)}{48 \pi \sin
(\pi(a/2-1))},\nonumber\\ f_{3}(a)&=&\frac{(2a-5)(2a-7)}{2 \sin (\pi
(a-3/2))}.\label{22}
\eea

After performing the integration of the corresponding Feynman integrals
in the renormalized two point correlation function $G^{(0,2)}$ we
obtain for the renormalization factor $Z_{\parallel}$ at one-loop order
(in the case $d=3$) the following result
\be
Z_{\parallel}=1+\frac{{\bar v}}{8}-\frac{\bar w}{8}\frac{2^{(a-3)/2}}{\cos
(\pi(3-a)/2)}(2^{(a-1)/2}-2^{(a+1)/2}-\frac{2^{(a-1)/2}}{a-5}).\label{Z11}
\ee

Combining the renormalization factor $Z_{\parallel}$ together with
one-loop pieces of the $\beta$-functions, according to Eq.(\ref{21}) we
finally obtain the following expression for the surface critical exponent
$\eta_{\parallel}^{sp}$

\be
\eta_{\parallel}^{sp}=-\frac{\bar v}{8} + \frac{\bar w}{8} \frac{2^{(a-3)/2}(4-a)}{\cos
(\pi(3-a)/2)}(2^{(a-1)/2}-2^{(a+1)/2}-\frac{2^{(a-1)/2}}{a-5}).
\ee

 The above equation should be calculated
at the long-range (LR) stable fixed point obtained in
\cite{Holovatch01},\cite{Prudnikov} up to two-loop approximation
 for different fixed values of the correlation parameter,
$2<a\leq 3$. The series for other surface critical exponents can be
calculated on the basis of surface scaling relations (see Appendix 2) and
one-loop series for bulk critical exponents
\bea
\nu^{-1}&=&2-\frac{\bar v}{4}+\frac{f_{1}(a)-f_{2}(a)}{2}{\bar
w},\nonumber\\
\eta&=&\frac{1}{2}f_{2}(a).\label{nueta}
\eea

The results of our calculation are presented in Table 1.
\begin{table}[htb]
\caption{\label{tab:tab1}Surface critical exponents of the
long-flexible polymer at the special transition $c=c_{ads}$ for
$d=3$ up to one-loop order calculated at the pure (the case $a=3$
with $(v^{*}=1.632,w^{*}=0)$) and LR stable fixed point for
different fixed values of the correlation parameter, $2<a< 3$.}
\begin{center}
\begin{tabular}{rrrrrrrrrr}
\hline $  a  $~&~$  v^{*}  $~&~$  w^{*}  $~&~$  \eta_{\parallel}
$~&~$  \eta_{\perp}  $~&~$  \beta_{1}  $~&~$  \gamma_{11}
$~&~$  \gamma_{1}  $~&~$  \delta_{1}  $~&~$  \delta_{11}  $ \\
\hline
 3.0  &  1.63  &  0.00  &  -0.204  &  -0.102  &  0.250  &  0.704  &  1.255  &
 6.020  &  3.816 \\

 2.9  &  4.13  &  1.47  &  -0.716  &  -0.358  &  0.154  &  1.024  &
 1.511  &  8.581  &  5.865  \\

 2.8  &  4.73  &  1.68  &  -0.843  &  -0.421  &  0.134  &  1.111  &  1.590  &
 9.213  &  6.372 \\

 2.7  &  5.31  &  1.81  &  -0.967  &  -0.483  &  0.117  &  1.200  &  1.675  &
 9.832  &  6.867 \\

 2.6  &  5.89  &  1.87  &  -1.094  &  -0.546  &  0.100  &  1.294  &
 1.768  &  10.467  &  7.376 \\

 2.5  &  6.48  &  1.89  &  -1.235  &  -0.616  &  0.081  &  1.398  &
 1.869  &  11.173  &  7.941  \\

 2.4  &  7.10  &  1.87  &  -1.404  &  -0.700  &  0.058  &  1.520  &  1.985  &
 12.017  &  8.617 \\

 2.3  &  7.76  &  1.84  &  -1.638  &  -0.816  &  0.019  &  1.676  &  2.122  &
 13.183  &  9.551 \\

\end{tabular}
\end{center}
\end{table}

As it is easy to see, in the case $a=d=3$, which corresponds to
random uncorrelated  pointlike disorder (or short-range-correlated
disorder) the obtained one-loop results for the surface critical
exponents conside with results for pure model (see
\cite{DSh98,Sh97} Pad\'e aproximants [1/0]). Besides, for the case
of a medium with long-range-correlated quenched disorder, the
process of adsorption of long-flexible polymer chains on a wall is
characterized by the new class of surface critical exponents. They
allow us to find useful characteristics which describe the
critical behaviour of long-flexible polymer chains at the
adsorption on a wall.

Since the bulk translational invariance is broken near the
surface, the number of configurations $Z_{N}$ depends on the
distance $z_{A}=z$ of the starting point of the chain from the
surface and behaves as $Z_{N}(z)\sim q^{N} N^{\gamma_{1}-1}$ near
the surface. A similar dependence on $z$ will be found for the
number of configurations $Z_{N}(z,z')$, when both ends of the
chain are at the surface (or close to it), i.e., $Z_{N}(z,z')\sim
q^{N} N^{\gamma_{11}-1}$. Another quantities of interest are $
Z(0,z')\sim z'^{a_{\lambda}} N^{b_{\lambda}} $ with
$a_{\lambda}=\eta_{\parallel}-\eta_{\perp}$,
$b_{\lambda}=-1+\gamma_{11}$ for $z'\ll lN^{\nu}$ ($l$ is the
effective monomer linear dimension, $l\to 0$ and $N\to \infty$)
and the value of the mean number of polymer chain free ends in a
layer between distances $z$ and $z+dz$ from the surface $(z=0)$ $
Z(z) \sim z^{-|a^{'}|} N^{b^{'}}$ for $z\ll lN^{\nu}$, where the
exponent $|a^{'}|$ characterizes density profiles for polymer
chain ends and equals $ |a^{'}|=(\gamma-\gamma_{1})/ \nu =
-\eta_{1}/2. $ The obtained results indicate that the presence of
long-range correlated quenched disorder facilitates the process of
adsorption of polymer chain on a hard wall. Besides, in a medium
with long-range correlated quenched disorder the swelling of the
polymer coil is governed by the exponent $\nu_{LR}$ that increases
when the correlation of the disorder increases (i.e. $a$
decreases). When $a<a_{marg}< 2.3$, then a crossover to the
collapse of the polymer is predicted. In the case of very strong
correlation of the disorder we have to deal with a situation when
such a polymer collapses and then adsorbs on a wall. The critical
exponents which characterize the process of adsorption of
long-flexible polymer chains inserted into the medium with
long-range correlated disorder belong to the new universality
class, different from that of the pure model.

\renewcommand{\theequation}{A1.\arabic{equation}}
\section*{Appendix 1}
\setcounter{equation}{0}
Applying the field-theoretical renormalization group (RG) approach we
perform calculations in a double expansion in $\epsilon=4-d$ and in
$\delta=4-a$ up to the one-loop order, as was proposed by Weinrib and
Halperin \cite{WH} for infinite systems. Thus,
for the renormalization factor $Z_{\parallel}$ we obtain
\bea Z_{\parallel}&=&1+\frac{{\bar v}}{4
(1+\epsilon)}-\frac{{\bar w}2^{-1-\delta}\sqrt{\pi}}{\cos(\frac{\pi}{2}
(\delta-\epsilon))\Gamma({\epsilon/2})}\left(\frac{\Gamma(\frac{\epsilon+\delta}{2})}
{\Gamma(\frac{1+\delta}{2})(1+\delta)}\right.+\nonumber\\
&+&\left.
\frac{(\delta-\epsilon)\Gamma(\frac{\epsilon+\delta}{2}-1)}{2\Gamma
(\frac{1+\delta}{2})}-\frac{\Gamma|\frac{\epsilon+\delta}{2}-2|}
{\Gamma|\frac{\delta-1}{2}|}\right).\label{eps1}
\eea
Combining the renormalization factor $Z_{\parallel}$ together with the
one-loop pieces of the beta functions, derived in
\cite{Holovatch01,Blavatska01} we obtain for the surface critical exponent
$\eta_{\parallel}^{sp}$
\bea
\eta_{\parallel}^{sp}&=&-\frac{{\bar
v}}{4}\frac{\epsilon}{(1+\epsilon)}+\frac{{\bar w}}{2}\frac{\delta
2^{-\delta}\sqrt{\pi}}{\Gamma(\frac{\epsilon}{2})\cos(\frac{\pi}{2}(\delta-\epsilon))}
\left(\frac{\Gamma(\frac{\epsilon+\delta}{2})}{\Gamma(\frac{1+\delta}{2})(1+\delta)}\right.+\nonumber\\
&+&\left.\frac{(\delta-\epsilon)\Gamma(\frac{\epsilon+\delta}{2}-1)}{2\Gamma(\frac{1+\delta}{2})}-
\frac{\Gamma|\frac{\epsilon+\delta}{2}-2|}{\Gamma|\frac{\delta-1}{2}|}\right).\label{eps2}
\eea

\renewcommand{\theequation}{A2.\arabic{equation}}
\section*{Appendix 2}
\setcounter{equation}{0}

The individual RG series expansions for the other critical exponents can
be derived through standard surface scaling relations \cite{D86} with
$d=3$ \begin{eqnarray}
&& \eta_{\perp} = \frac{\eta +
\eta_{\parallel}}{2}, \nonumber\\
&& \beta_{1} = \frac{\nu}{2}
(d-2+\eta_{\parallel}), \nonumber\\
&& \gamma_{11}=\nu(1-\eta_{\parallel}), \nonumber\\
&& \gamma_{1}= \nu(2-\eta_{\perp}), \label{sc}\\
&& \Delta_{1}= \frac{\nu}{2} (d-\eta_{\parallel}), \nonumber\\
&& \delta_{1} = \frac{\Delta}{\beta_{1}} =
\frac{d+2-\eta}{d-2+\eta_{\parallel}}, \nonumber\\
&& \delta_{11} = \frac{\Delta_{1}}{\beta_{1}}=
\frac{d-\eta_{\parallel}}{d-2+\eta_{\parallel}}\;.\nonumber
\end{eqnarray}

Each of these critical exponents characterizes certain properties of
the semi-infinite systems with long-range quenched disorder, in the
vicinity of the critical point. The values $\nu$, $\eta$, and
$\Delta=\nu(d+2-\eta)/2$ are the standart bulk exponents.

\section*{Acknowledgments}
 This work was supported by NATO Science Fellowships National
Administration under Grant No. 14/B/02/PL.

\end{document}